\documentclass{webofc}
\usepackage[varg]{txfonts}   
\usepackage{color}
\begin{document}
\title{Geometrical scaling of prompt
photons in heavy ion collisions}
%
%

\author{\firstname{Micha{\l}} \lastname{Prasza{\l}owicz}\inst{1}
\fnsep\thanks{\email{michal@if.uj.edu.pl}}}

\institute{M. Smoluchowski Institute of Physics, Jagiellonian University,  S. {\L}ojasiewicza 11,
30-348 Krak{\'o}w, Poland}

\abstract{%
We discuss geometrical scaling (GS) for the prompt (direct)  photons produced in heavy ion collisions. To this end we first introduce
the concept of  GS and illustrate its emergence on the example of charged particles. Next, we analyse direct photon data from
RHIC and from the LHC. We show that the data support the  hypothesis of GS in terms of participant number related to different centrality
classes. We also study GS at different energies, however a more detailed study will be possible only when data for at least three different energies
from one experiment will be available and also when we use the data obtained from the collisions of large systems (Cu+Cu, Au+Au, Pb+Pb) and small systems (p+p, d+Au, p+Au).
}
\maketitle
\section{Introduction}
\label{intro}
It is now widely accepted that in high energy scattering the bulk features of the produced
particles spectra give access to the properties of the initial state. Especially, in the small
Bjorken $x$ kinematical region, where the initial hadrons consist predominantly from the overoccupied 
gluonic cloud, a phenomenon called {\em geometrical scaling} (GS)  arrises. Geometrical scaling has been
first observed in the inclusive deep inelastic electron-proton scattering (DIS)~\cite{Stasto:2000er}, where the reduced cross section,
essentially $F_2(x,Q^2)/Q^2$, which is in principle a function of two variables $x$ and $Q^2$, depends in
fact on only one scaling variable $\tau=Q^2/Q^2_{\rm sat}(x)$. The saturation scale
\begin{equation}
Q^2_{\rm sat}(x)=Q_0^2 \left( x/x_0 \right)^{-\lambda}
\label{Qsat}
\end{equation}
is directly proportional to the gluon density in the proton, and $x_0$ and $Q_0$ are fixed parameters of the order
of  $10^{-3}$ and 1 GeV/$c$, respectively. Exponent $\lambda \approx 0.33$ is a nonperturbative dynamical 
quantity following from the properties of
the (non-linear) QCD evolution equation, however its numerical value has to be fixed from the data
(for review see Ref.~\cite{McLerran:2010ub}).

The properties of the saturated gluon densities should have an impact on the particles produced  in  proton-proton collisions.
Indeed, adopting a so called parton-hadron duality~\cite{Dokshitzer:1991eq}, one can argue that the bulk properties of the gluon spectra are shared by
charged particles that are eventually detected experimentally. In particular they exhibit GS in the small $x$ region
({\em i.e.} large energy $\sqrt{s}$ and moderate $p_{\rm T}$) \cite{McLerran:2010ex,McLerran:2010wm}. However,
multiplicity spectra scale with a somewhat lower power $\lambda$ than the one extracted from DIS. We have argued
that for pp scattering, where one does not control the overlap area transverse to the center of mass  reaction axis, the
quantity that scales with $\lambda \approx 0.33$ is the differential cross section, rather than multiplicity~\cite{Praszalowicz:2015dta}. This
should be contrasted with the heavy ion (HI) reactions, when the transverse overlap area is controlled by an appropriate
choice of centrality classes. 

The fact that GS exhibited by the  gluons  is transferred to the final state particles, which are created in the
nonperturbative hadronization process, which undergo final state interactions, which are being produced by resonance decays,
is by no means obvious. This is even more so in the case of HI collisions where the quark-gluon plasma (QGP) is created and
undergoes hydrodynamical evolution. Nevertheless, following Refs.~\cite{Praszalowicz:2015vba,Praszalowicz:2015hia},  
we will show in the next 
section that GS  in charged particles
spectra is indeed present even in this case.

From this point of view direct photons (by definition photons that {\underline{do} \underline{not}} originate from
hadronic decays) are an excellent probe of the initial state of HI collisions since they do not interact strongly while passing
through the quark-gluon plasma. However, photons are produced from quarks (through annihilation and Compton scattering), and therefore
they do not probe the overoccupied gluonic cloud directly. To this end one employs the Color Glass Condensate (CGC)
effective theory where the initial gluonic CGC evolves into an intermediate state called glasma, which is a strongly interacting
not thermalised QGP~\cite{McLerran:2010ub}. Quarks are produced in the thermalisation process, 
and -- if there are no other mass scales around -- their
distribution should exhibit geometrical scaling\footnote{Detailed description
of the photon production mechanism in glasma is beyond scope of this report; we refer the reader to 
Refs.~\cite{Chiu:2012ij,McLerran:2015mda,Khachatryan:2018ori}.} 
in variable $\tau = p_{\rm T}^2/Q^2_{\rm sat}(x)$. Similarly, the photon spectra
should scale as well. In order to validate this scenario one should check if the available photon data exhibit geometrical scaling. The first
step in this direction has been undertaken in Ref.~\cite{Klein-Bosing:2014uaa} where the functional form of the photon spectra 
has been assumed to be of the form 
$p_{\rm T}^{-n}$ (see also Refs.~\cite{Chiu:2012ij,McLerran:2015mda,Khachatryan:2018ori}).
Here, following 
Refs.~\cite{McLerran:2010ex,McLerran:2010wm,Praszalowicz:2015dta,Praszalowicz:2015vba,Praszalowicz:2015hia}, 
we shall employ model independent  {\em method of ratios} to analyse GS of photon spectra, 
which is done in Sect.~\ref{photons}. But first, in Sect.~\ref{charged}, we shortly recall analysis of charged particle spectra already
reported in Refs.~\cite{Praszalowicz:2015vba,Praszalowicz:2015hia}. We conclude in Sect.~\ref{concl}.

\section{Charged particles}
\label{charged}

By geometrical scaling
\cite{Stasto:2000er} of charged particles
in hadronic collisions 
we mean that the multiplicity distributions are  well described -- up to the logarithmic correction
of the running coupling constant -- by a universal function $F(\tau)$ 
\cite{McLerran:2010ex,McLerran:2010wm}:
\begin{equation}
\frac{1}{S_{\bot}}\frac{dN_{\text{ch}}}{d\eta d^{2}p_{\text{T}}}%
=\,F(\tau)\label{Npp}%
\end{equation}
of the scaling variable%
\begin{equation}
\tau=\frac{p_{\text{T}}^{2}}{Q_{\text{sat}}^{2}}.\label{taupp}%
\end{equation}
Saturation scale (for particles produced in mid rapidity region) reads therefore%
\begin{equation}
Q_{\text{sat}}^{2}=Q_{0}^{2}\left(  \frac{p_{\text{T}}}{Wx_{0}}\right)
^{-\lambda}.\label{Qsatpp}%
\end{equation}
Here $W=\sqrt{s}$ is the scattering energy and for the reference we take
$x_{0}=10^{-3}$ and $Q_{0}=1$ GeV/$c$. Parameter $S_{\bot}$ is a transverse
area, which for heavy ion collisions
corresponds to geometrical overlap of the colliding nuclei at given impact
parameter $b$ \cite{Kharzeev:2004if}.
HI data are usually divided into centrality classes
that select events within certain range of impact parameter $b$. In this case
both transverse area $S_{\bot}$ and the saturation scale $Q_{\text{sat}}^{2}$
acquire additional dependence on centrality that is characterized by an
average number of participants $N_{\text{part}}$. 
We have \cite{Klein-Bosing:2014uaa,Kharzeev:2004if}:%
\begin{equation}
S_{\bot}\sim N_{\text{part}}^{\delta}\;\; \text{and} \;\; Q_{\text{sat}}^{2}\sim
N_{\text{part}}^{\delta/2}.\label{Npartscaling}%
\end{equation}
where one typically assumes $\delta=2/3$, which follows from the collision geometry. Therefore in HI collisions%
\begin{equation}
\frac{1}{N_{\text{evt}}}\frac
{dN_{\text{ch}}}{N_{\text{part}}^{\delta} \, 2\pi p_{\text{T}}\,d\eta dp_{\text{T}}}=\,
\frac{1}{Q_{0}^{2}} \, F(\tau)\label{multHI}%
\end{equation}
and the scaling variable $\tau$ takes the following form:%
\begin{equation}
\tau=\frac{p_{\text{T}}^{2}}{N_{\text{part}}^{\delta/2} \, Q_{0}^{2}}\left(
\frac{p_{\text{T}}}{W}\right)  ^{\lambda}.\label{tauHI}%
\end{equation}
Taking
$x_{0}=10^{-3}$ we have that the  energy $W$ should be expressed in TeV, whereas $p_{\text{T}}$
in GeV. The analysis presented here will be only qualitative and we shall not fine tune the
scaling exponent $\lambda$ keeping it fixed at $\lambda=0.3$.

In this section we shall analyse  ALICE data on PbPb collisions at 2.76 TeV
\cite{Abelev:2012hxa}
and also earlier data from RHIC from STAR \cite{Adams:2003kv,Adler:2002xw}
and PHENIX \cite{Adler:2003au,Adcox:2001jp}
collaborations at 200 and
130 GeV per nucleon respectively. Centrality classes together with participants numbers
are given in Table~\ref{tab:cent}.

\begin{table}[h!]
\caption{Centrality classes and the corresponding numbers of participants in heavy ion experiments
analysed in this paper. Energies per nucleon in TeV are displayed next to the experiment name.
Bold face entries in blue (color on-line) show classes of similar number of participants analysed in the text.}
\label{tab:cent}
\centering
\begin{tabular}
[c]{|rr|rr|rr|rr|}\hline
\multicolumn{2}{|c|}{{ALICE 2.76 }} & \multicolumn{2}{|c|}{{STAR 0.2
\& 0.13 }} & \multicolumn{2}{|c|}{{PHENIX 0.2 }} &
\multicolumn{2}{|c|}{{PHENIX 0.13 }}\\
{centrality} & $N_{\text{part}}$ & {centrality} & $N_{\text{part}}$ &
{centrality} & $N_{\text{part}}$ & {centrality} & $N_{\text{part}}$ \\
\hline
0-5\% & 383 &  &  &  &  &  & \\
\textcolor{blue}{\bf 5-10\%} & \textcolor{blue}{\bf 330} & \textcolor{blue}{\bf 0-5\%} &
\textcolor{blue}{\bf 350} & \textcolor{blue}{\bf 0-10\%} & \textcolor{blue}{\bf 352.2} &
\textcolor{blue}{\bf 0-5\% } & \textcolor{blue}{\bf 348}\\
&  & 5-10\% & 296 &  &  & 5-15\% & 271\\
10-20\% & 261 & 10-20\% & 232 & 10-20\% & 234.6 &  & \\
\textcolor{blue}{\bf 20-30\%} & \textcolor{blue}{\bf 186} &
\textcolor{blue}{\bf 20-30\%} & \textcolor{blue}{\bf 165} &
\textcolor{blue}{\bf 20-30\%} & \textcolor{blue}{\bf 166.6} &
\textcolor{blue}{\bf 15-30\%} & \textcolor{blue}{\bf 180}\\
30-40\% & 129 & 30-40\% & 115 & 30-40\% & 114.2 &  & \\
40-50\% & 85 &  &  & 40-50\% & 74.4 & 30-60\% & 79\\
50-60\% & 53 & 40-60\% & 62 & 50-60\% & 45.5 &  & \\
\textcolor{blue}{\bf 60-70\%} & \textcolor{blue}{\bf 30} & \textcolor{blue}{\bf 60-80\%} &
\textcolor{blue}{\bf 20} & \textcolor{blue}{\bf 60-70\%} & \textcolor{blue}{\bf 25.7} &
\textcolor{blue}{\bf 60-80\%} & \textcolor{blue}{\bf 19.5}\\
70-80\% & 15.8 &  &  & 70-80\% & 13.4 &  & \\
&  &  &  & 80-92\% & 6.3 & 80-92\% & 5.5\\\hline
\end{tabular}
\end{table}

\begin{figure}[h!]
\centering
\includegraphics[width=6.7cm]{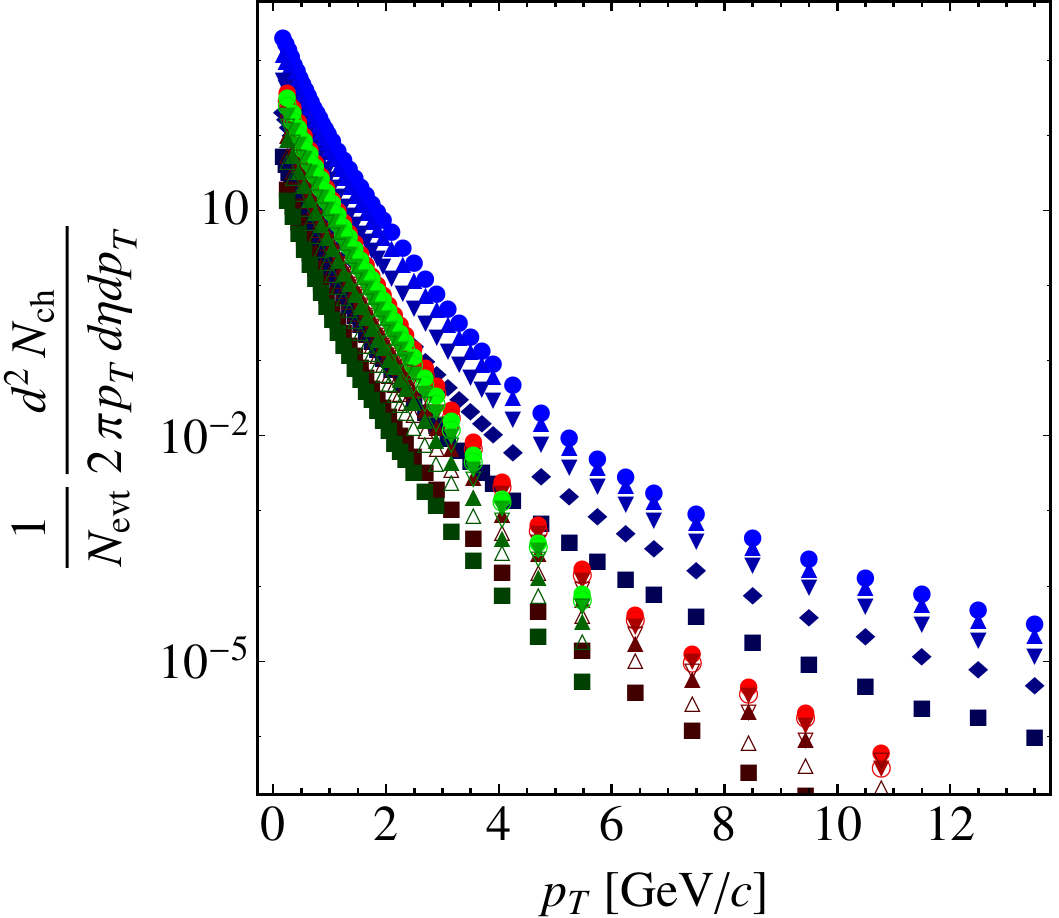}
\includegraphics[width=6.7cm]{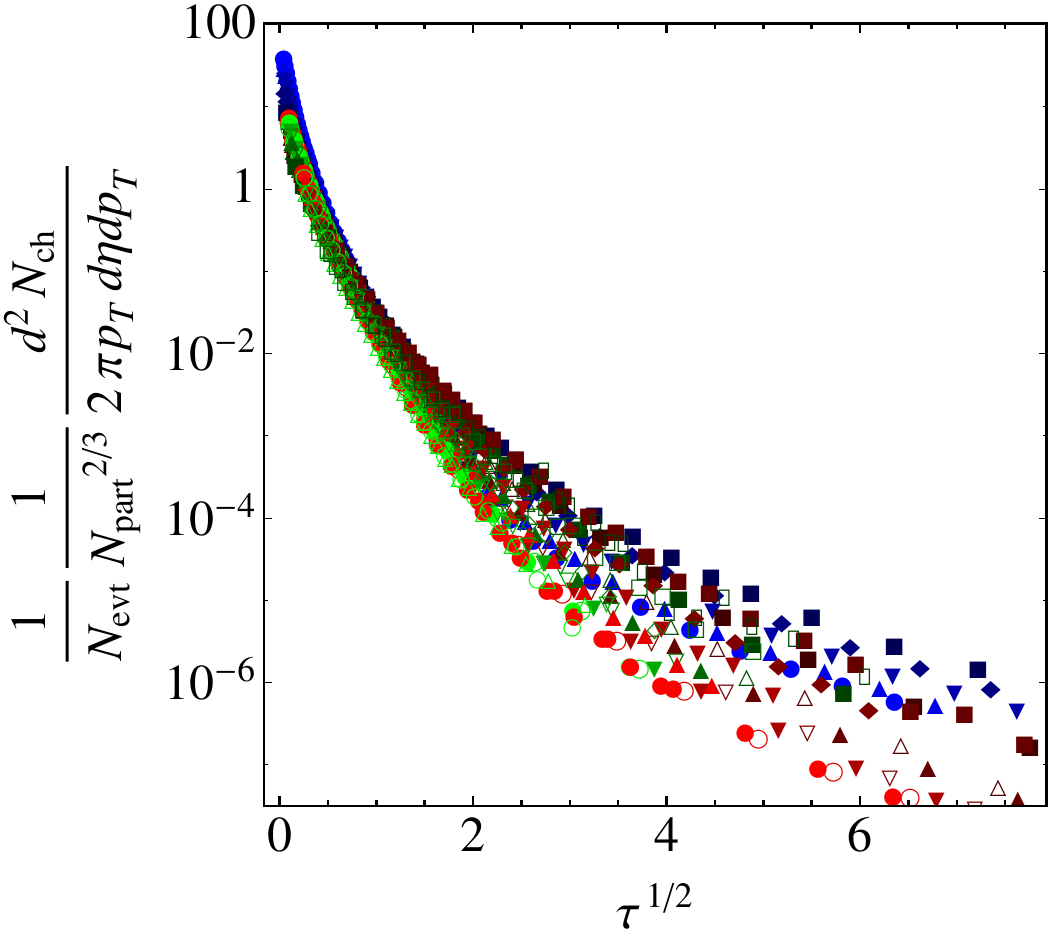}
\caption{Illustration of geometrical scaling in heavy ion collisions at different energies and
different centrality classes. Left panel shows charged particle distributions from 
ALICE \cite{Abelev:2012hxa}, 
STAR \cite{Adams:2003kv,Adler:2002xw} and 
PHENIX \cite{Adler:2003au,Adcox:2001jp}
plotted as functions of $p_{\rm T}$. In the right panel the same distributions are 
scaled according to Eq.~(\ref{multHI}). Symbols used here are the same as in 
Figs.~\ref{fig:midc}--\ref{fig:Alice}. }%
\label{fig:all}
\end{figure}

In Fig.~\ref{fig:all} we plot multiplicity distributions from all three experiments displayed in Table~\ref{tab:cent},
for different centrality classes and -- in the case of STAR and PHENIX -- for both scattering energies.
First in the left panel multiplicity distributions are plotted as functions of $p_{\rm T}$ and then in the right panel 
as functions of scaling variable $\sqrt{\tau}$ (\ref{tauHI}). We see that different
distributions from the left panel coincide over some range of  $\sqrt{\tau}$ when scaled according to Eq.~(\ref{multHI}).
For clarity in Fig.~\ref{fig:all} we have used only every second centrality class of ALICE and PHENIX@200~GeV data. 

We can see from Fig.~\ref{fig:all} that, indeed, heavy ion data scale according to (\ref{multHI})
up to $\sqrt{\tau}\sim 1.8$ approximately. The quality of scaling is, however, not as good as in
the case of pp scattering \cite{McLerran:2010wm}. One could perhaps improve the quality of GS
by tuning the exponent $\lambda$ in the definition of $\tau$ (\ref{tauHI}). We have decided to keep
$\lambda$ constant for the purpose of present analysis because of the systematic differences between
the data from different collaborations. The data has been taken in the rapidity intervals
that are different in different experiments, also the partition of the data into centrality classes varies
from one experiment to another as can be seen from Table~\ref{tab:cent}.

\begin{figure}[h!]
\centering
\includegraphics[width=6.7cm]{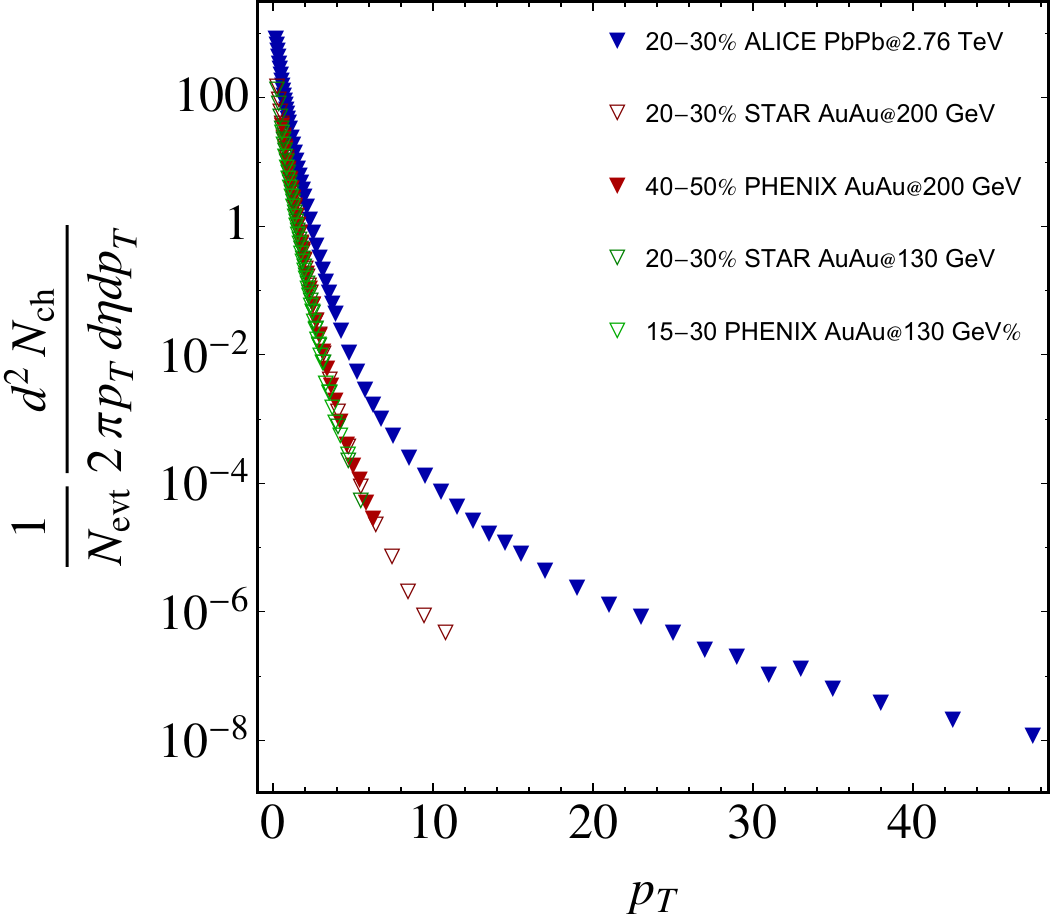}
\includegraphics[width=6.7cm]{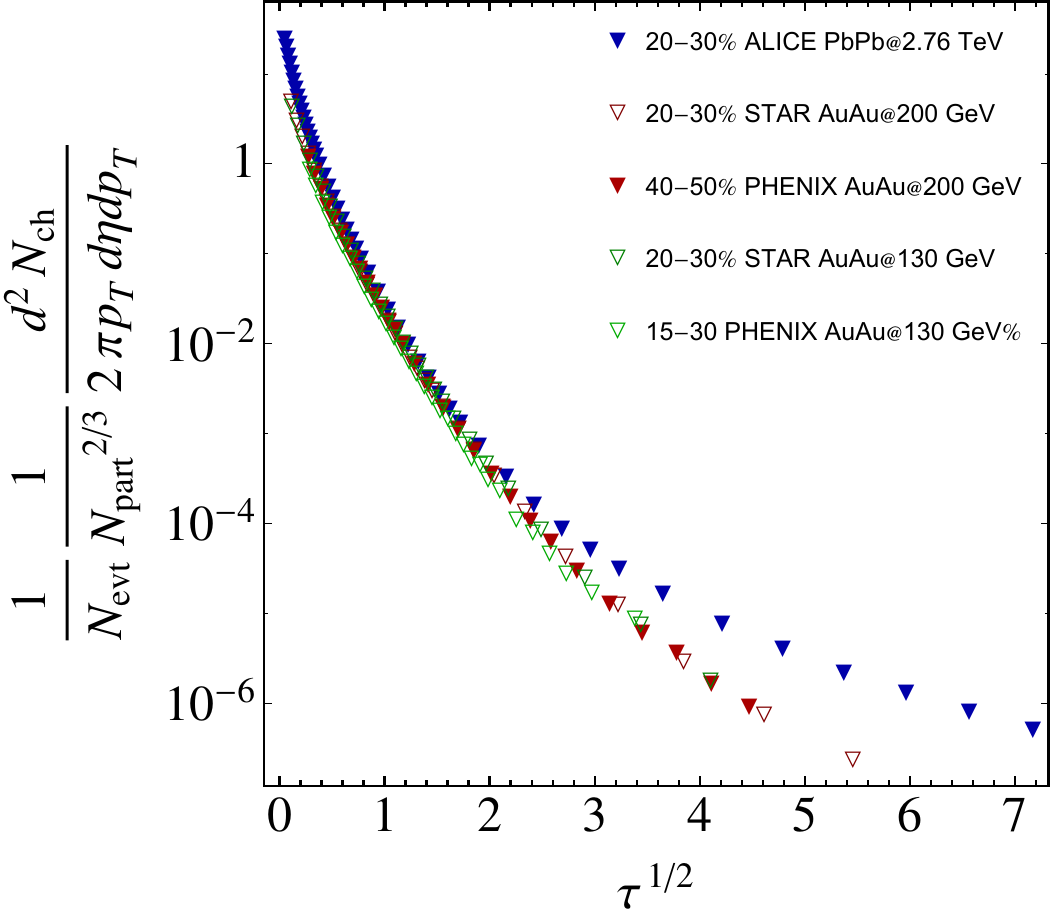}
\caption{Illustration of geometrical scaling in heavy ion collisions at different energies for
mid centrality classes corresponding to $n_{\rm part}= 165 - 186$. 
 Left panel shows charge particle distributions from 
ALICE \cite{Abelev:2012hxa}, 
STAR \cite{Adams:2003kv,Adler:2002xw} and 
PHENIX \cite{Adler:2003au,Adcox:2001jp}
 plotted as functions of $p_{\rm T}$. In the right panel the same distributions are 
scaled according to Eq.~(\ref{multHI}). }%
\label{fig:midc}
\end{figure}

It is interesting to test now the quality of GS separately in dependence on energy and on centrality ({\em i.e.} on $N_{\rm part}$).
Let us first discuss scaling with energy by selecting centrality classes that correspond to
similar number of participants in all three experiments. For illustration we plot in
 in Fig.~\ref{fig:midc} multiplicity distributions with $n_{\rm part}= 165 - 186$ 
(second blue row in Table~\ref{tab:cent}). We see rather good scaling
up to $\sqrt{\tau} \approx 2$. The plots for other centralities look very much the same.

\begin{figure}[h!]
\centering
\includegraphics[width=6.7cm]{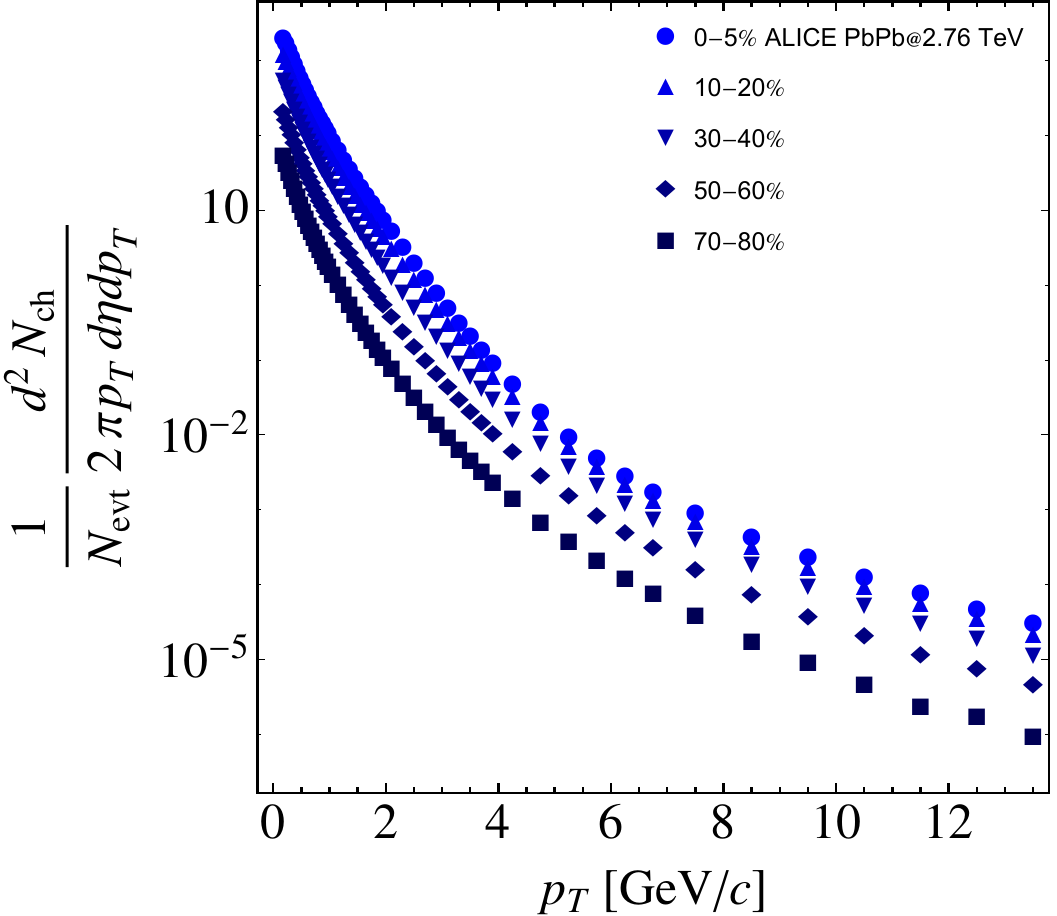}
\includegraphics[width=6.7cm]{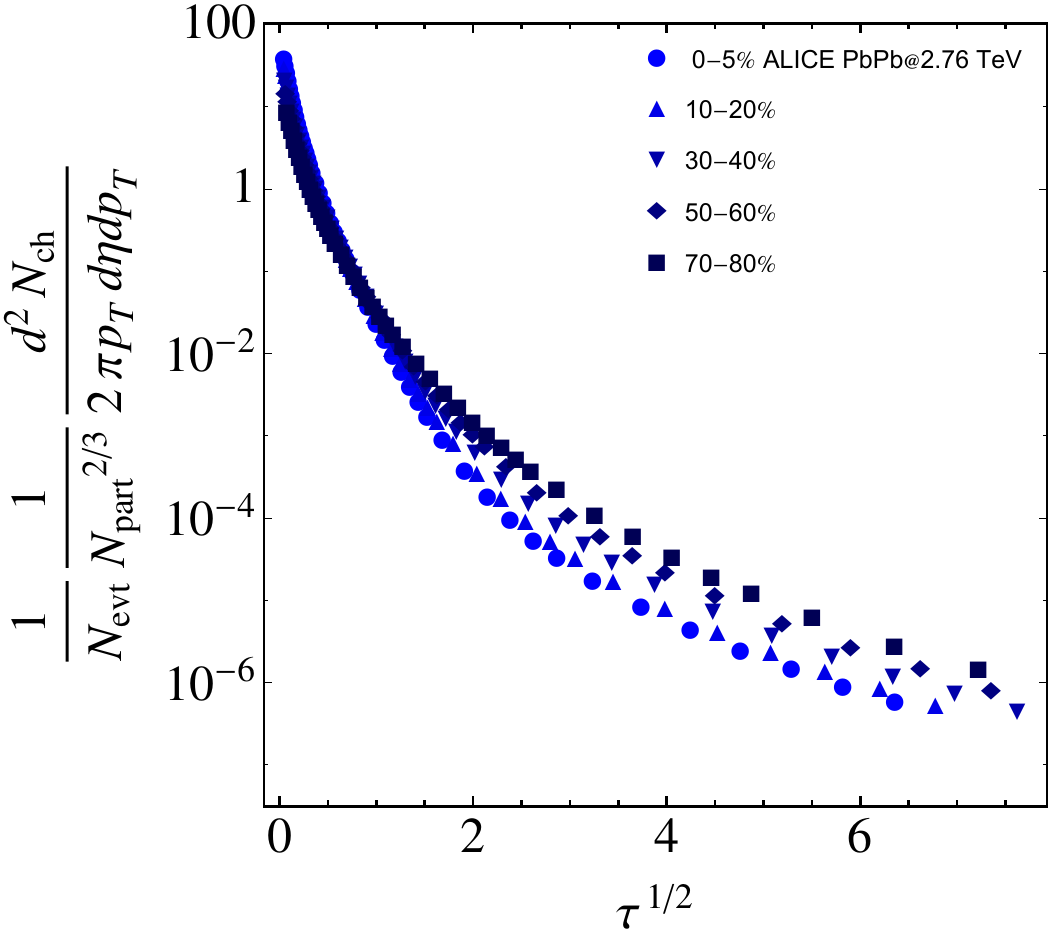}
\caption{Illustration of geometrical scaling in heavy ion collisions at fixed ALICE \cite{Abelev:2012hxa}
energy
of 2.76 TeV for different centrality classes.
 Left panel shows charge particle distributions plotted as functions of $p_{\rm T}$. 
 In the right panel the same distributions are 
scaled according to Eq.~(\ref{multHI}). }%
\label{fig:Alice}
\end{figure}

For fixed scattering energy $W$ equation~(\ref{multHI}) relates distributions of different 
participant number. We shall now examine the quality of this scaling by plotting multiplicity
distributions at the same energy but different centrality classes. In Fig.~\ref{fig:Alice}
we plot ALICE data at 2.76 GeV. Plots for RHIC energies are very similar, and one can say
that generally centrality scaling is of worse quality than the energy scaling discussed above.

\section{Photons}
\label{photons}

In this paper we shall analyse the following data sets on direct photons:
PHENIX \cite{Adare:2008ab,Adare:2014fwh} Au+Au~@~200~GeV with the following centrality classes
0-20\% ($N_{\rm part}=277.5$), 20-40\% ($N_{\rm part}=135.6$), 40-60\% ($N_{\rm part}=56.0$),
60-92\% ($N_{\rm part}=12.5$) and ALICE\footnote{Special thanks to Jacek Otwinowski for providing us with the pertinent
values of $N_{\rm part}$.} \cite{Wilde:2012wc,Adam:2015lda} Pb+Pb~@~2.76~TeV: 0-20\% ($N_{\rm part}=308$), 20-40\%  ($N_{\rm part}=157$)
and 40-80\% ($N_{\rm part}=45.7$). More recent PHENIX data~\cite{Adare:2018wgc}  reported this year has not been available
at the time of preparing this manuscript.

\subsection{$N_{\rm part}$ scaling}
\begin{figure}[h!]
\centering
\includegraphics[width=6.3cm]{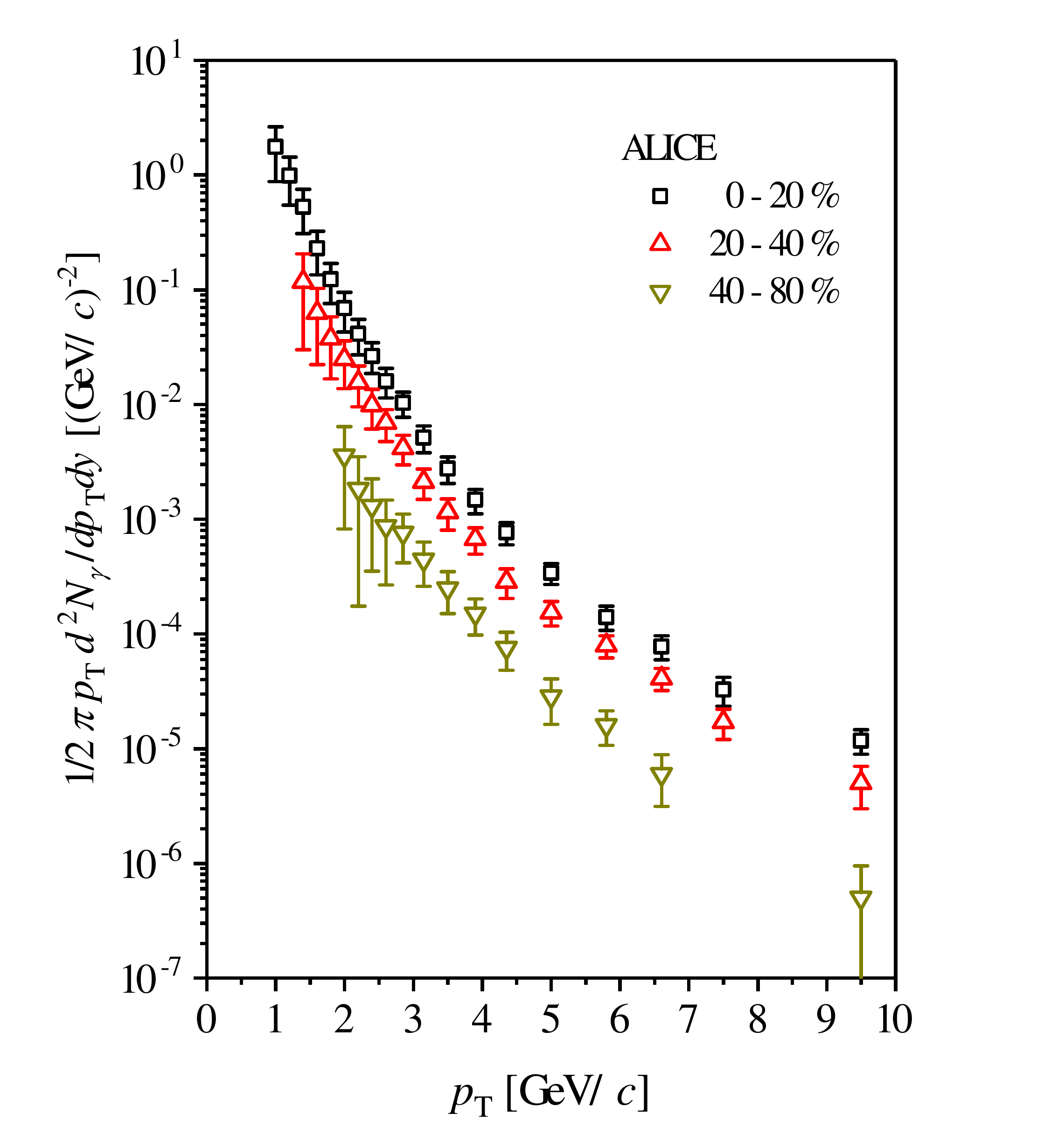}
\includegraphics[width=6.3cm]{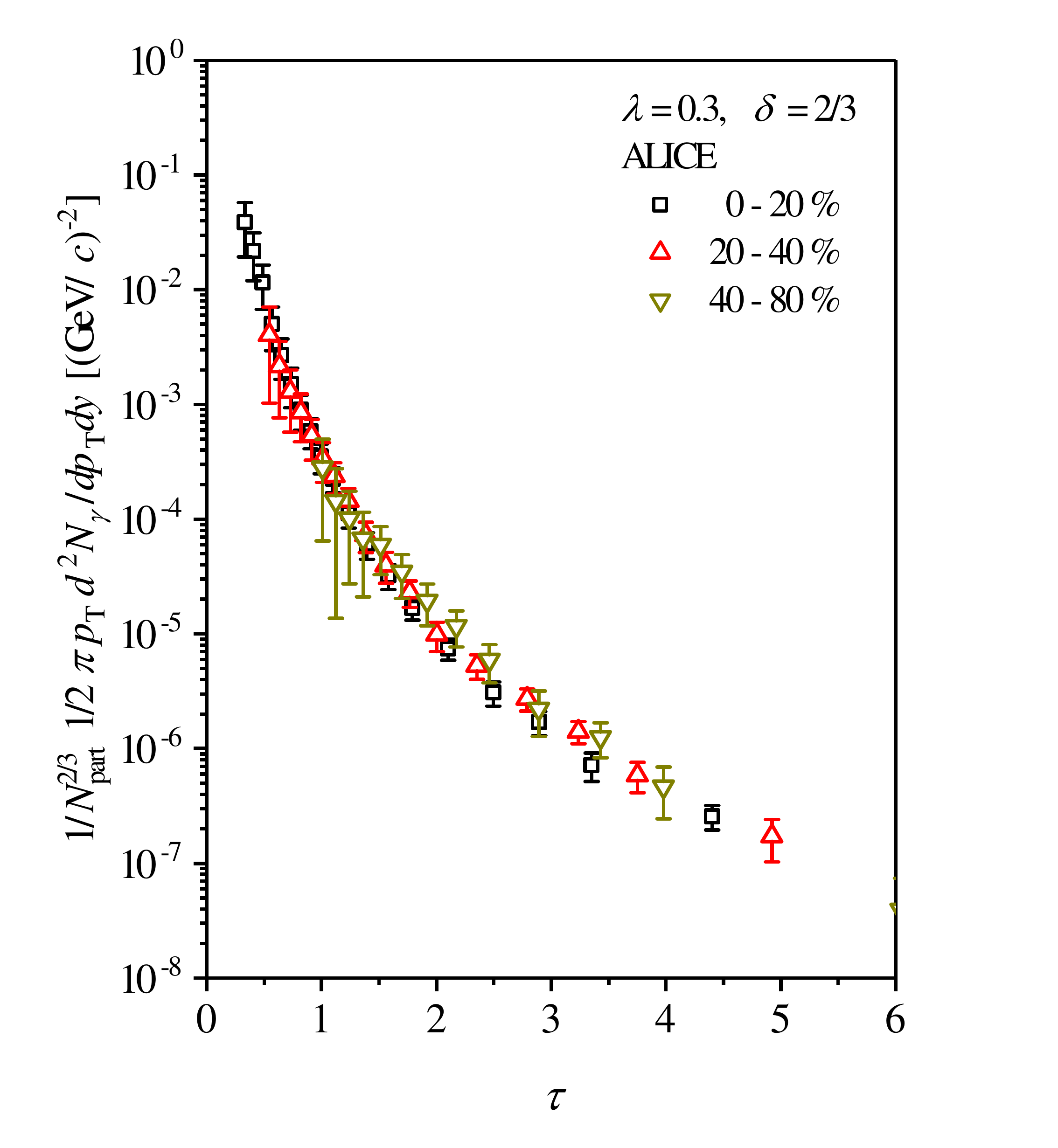}
\caption{Illustration of $N_{\rm part}$ geometrical scaling of the $\gamma$ yields in heavy ion collisions for
different centrality classes for ALICE PbPb data. Left panel: transverse momentum spectra of direct photons at three centrality classes:
0-20\% (black squares), 20-40\% (red up-triangles) 40-80\%, (dark green down-triangles). Right panel: scaled spectra for $\delta=2/3$.
Data from Ref.~\cite{Adam:2015lda}.}%
\label{fig:alice0}
\end{figure}

\begin{figure}[h!]
\centering
\includegraphics[width=6.3cm]{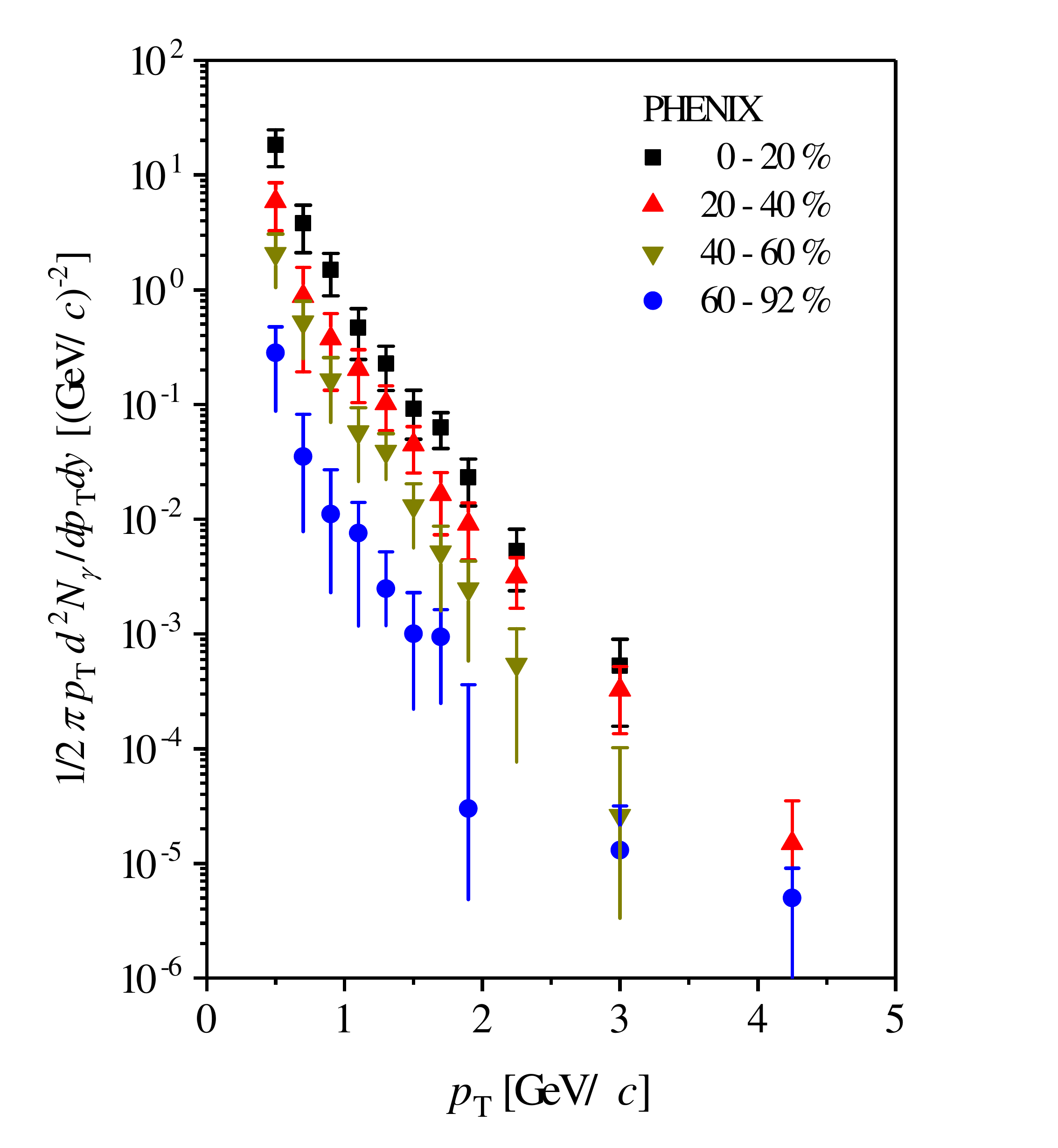}
\includegraphics[width=6.3cm]{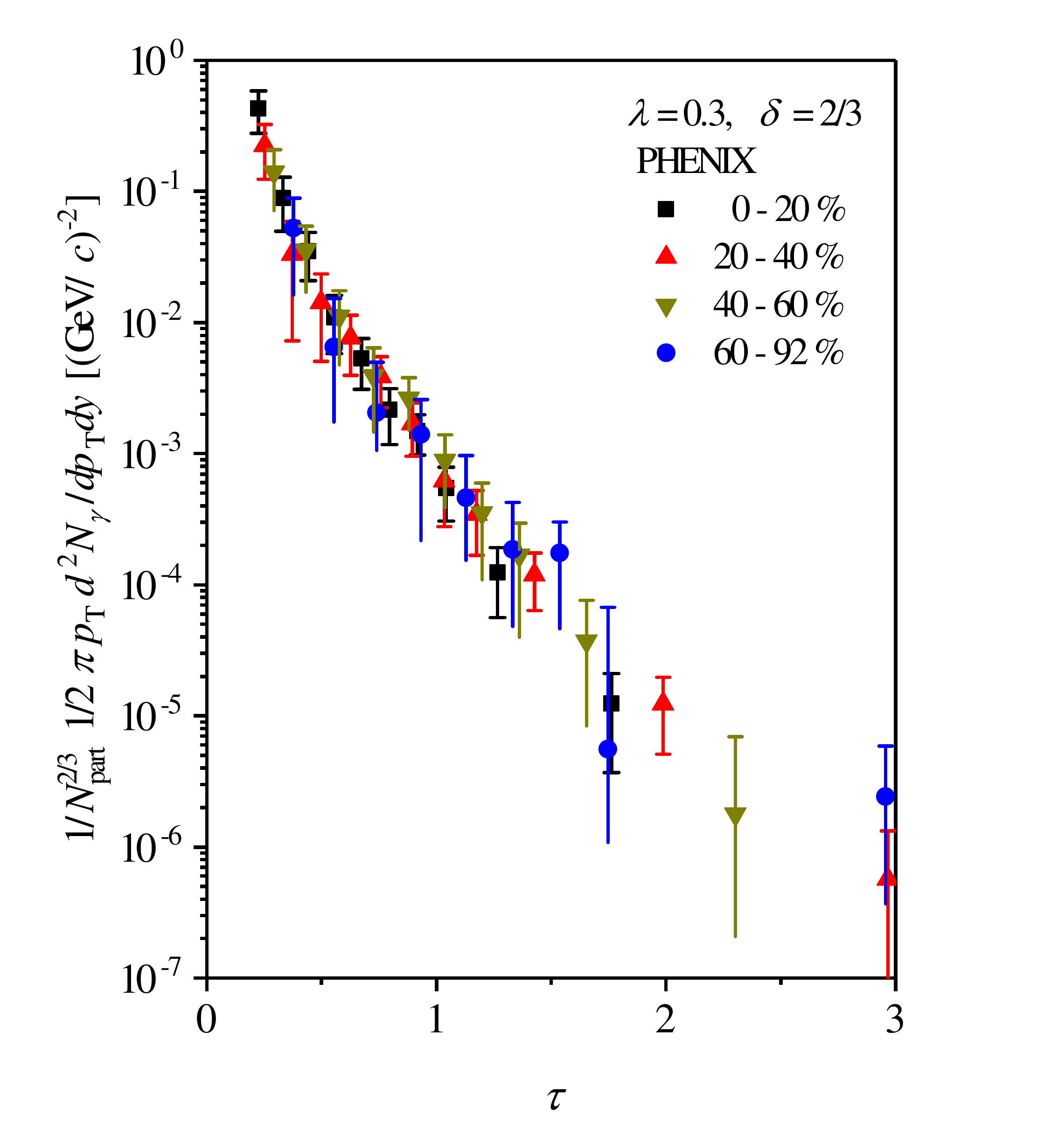}
\caption{Illustration of $N_{\rm part}$ geometrical scaling of the $\gamma$ yields in heavy ion collisions for
different centrality classes for PHENIX AuAu data at 200~GeV~\cite{Adare:2014fwh}. Left panel: transverse momentum spectra of direct photons at four 
centrality classes:
0-20\% (black squares), 20-40\% (red up-triangle) 40-80\% (dark green down-triangles),
6--92\% (blue circles). Right panel: scaled spectra for $\delta=2/3$.
Lower error bars without end caps have been arbitrarily shortened for better visibility.}%
\label{fig:phenix1}
\end{figure}

Let us first examine the $N_{\rm part}$ dependence of geometrical scaling for  the ALICE data \cite{Adam:2015lda}.
These spectra ale plotted in the left panel of Fig.~\ref{fig:alice0}, where we have included points with $p_{\rm T} \le 10$~GeV/$c$.
In the right panel the same spectra ale plotted after rescaling according to (\ref{multHI}) for $\delta=2/3$. We see that to a very good
accuracy all three spectra coincide. This result should be contrasted with the charged particle scaling shown in Fig~\ref{fig:Alice}, which is of much worse quality. The same analysis is illustrated in Fig.~\ref{fig:phenix1} for PHENIX data \cite{Adare:2008ab,Adare:2014fwh}.

\begin{figure}[h]
\centering
\includegraphics[width=6.8cm]{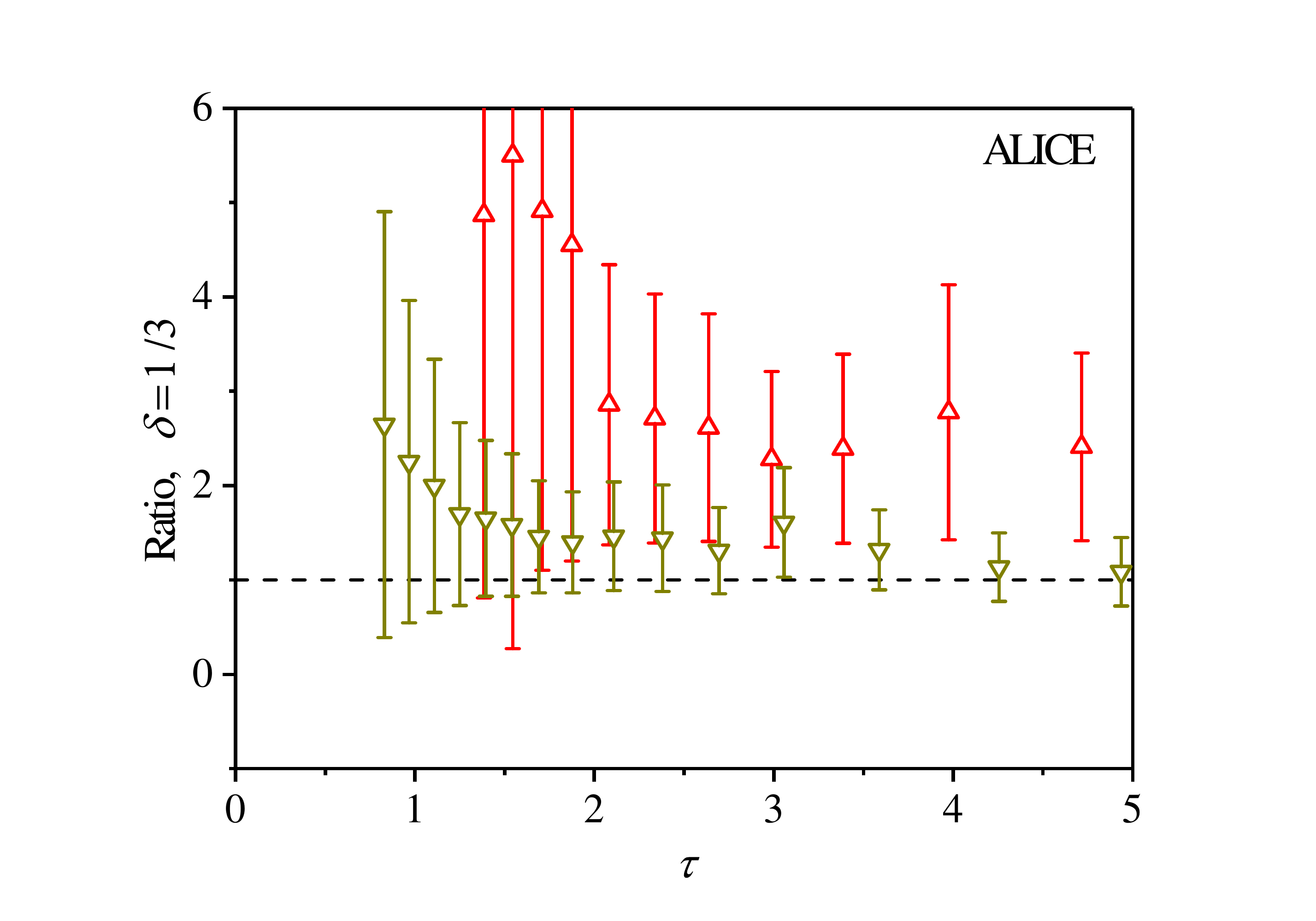}~\includegraphics[width=6.5cm]{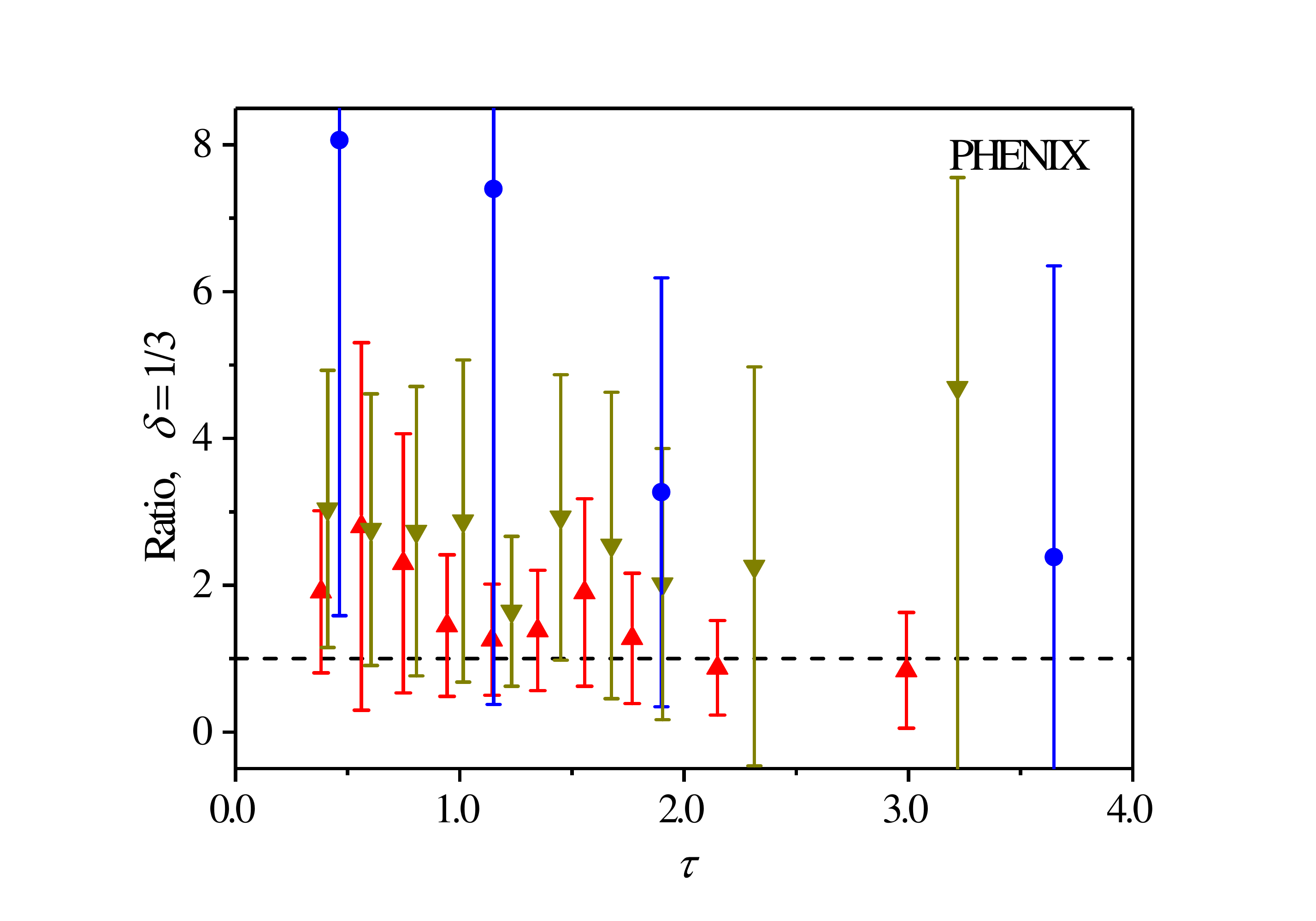}\\
\includegraphics[width=6.8cm]{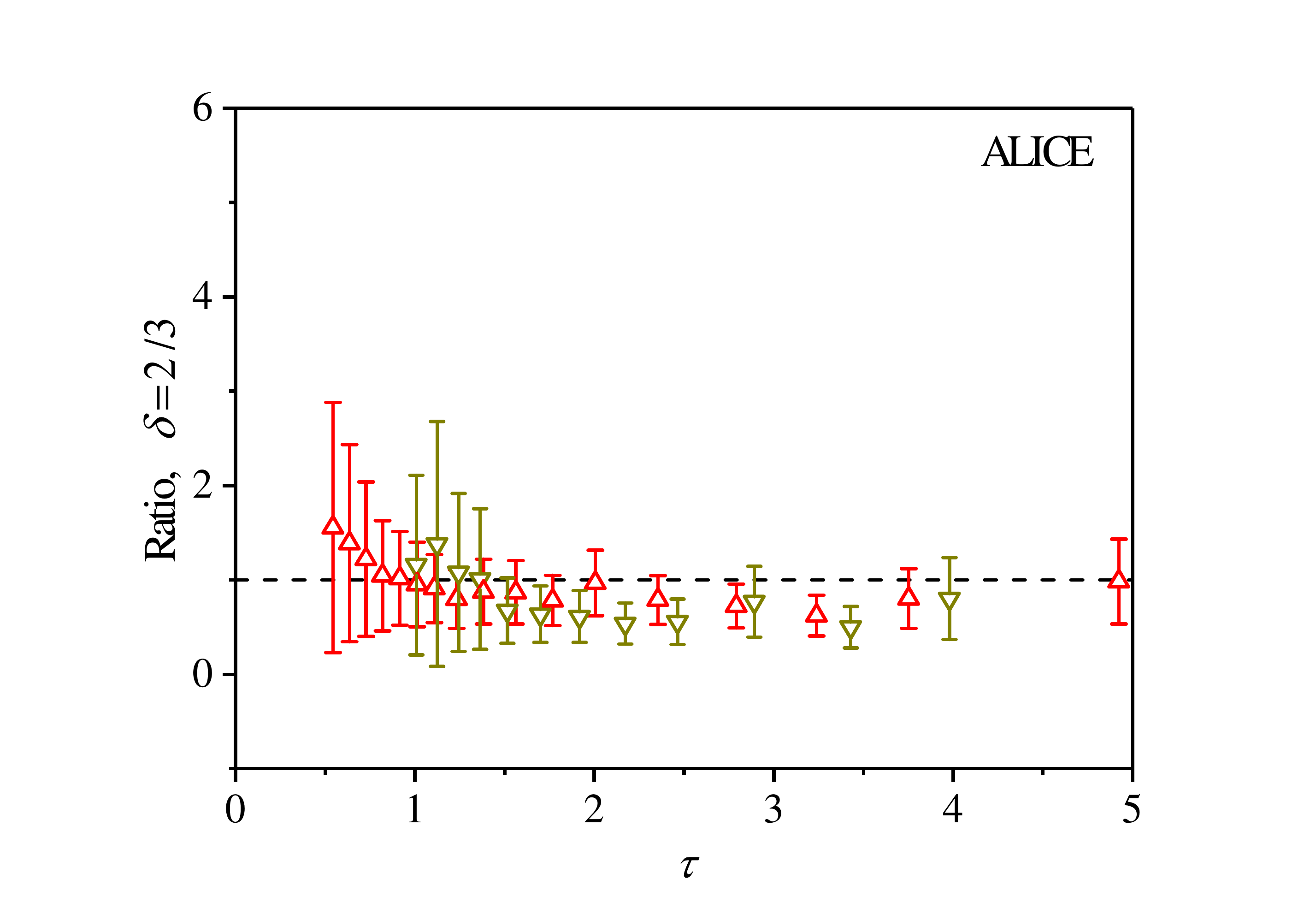}~\includegraphics[width=6.5cm]{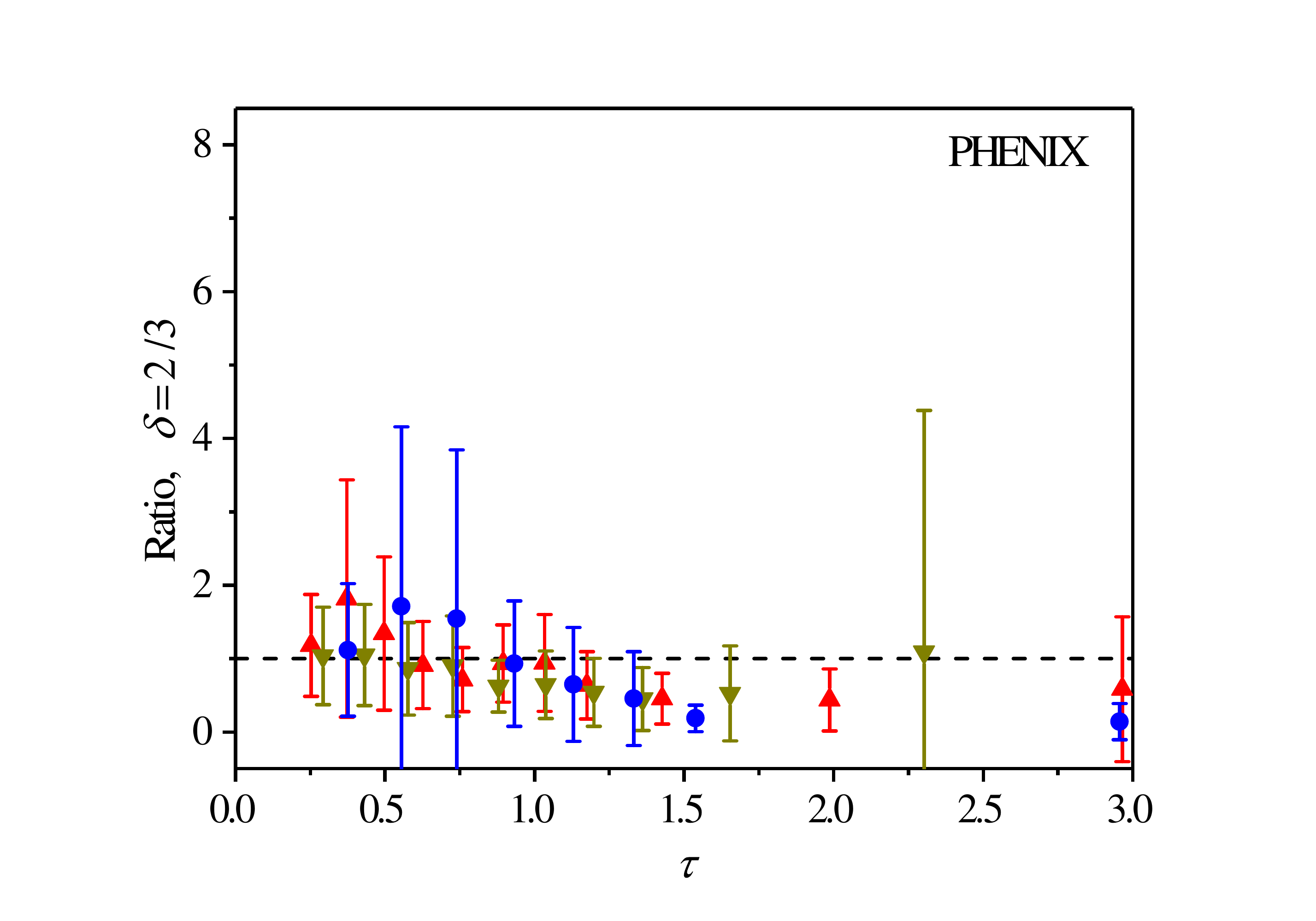}\\
\includegraphics[width=6.8cm]{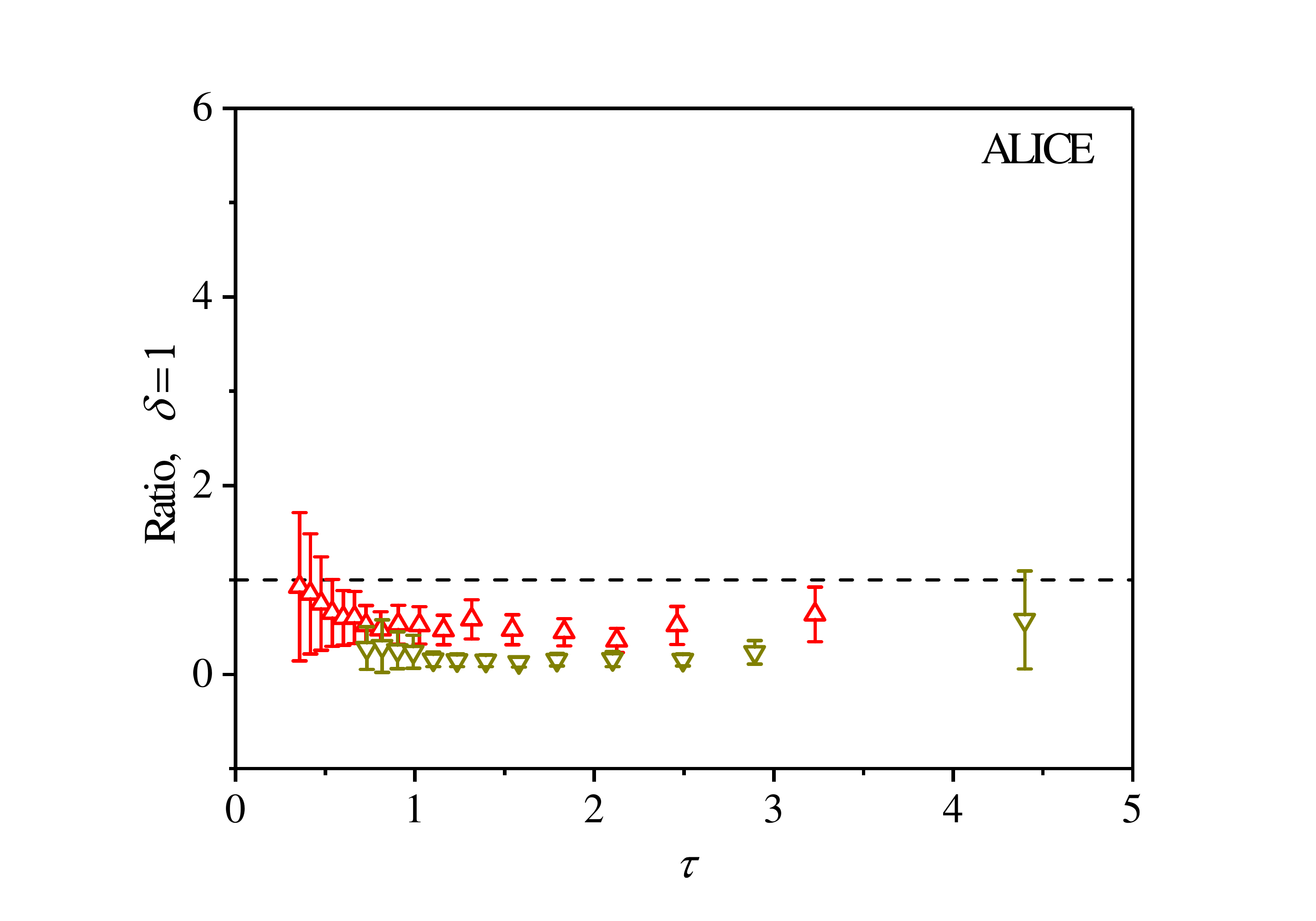}~\includegraphics[width=6.5cm]{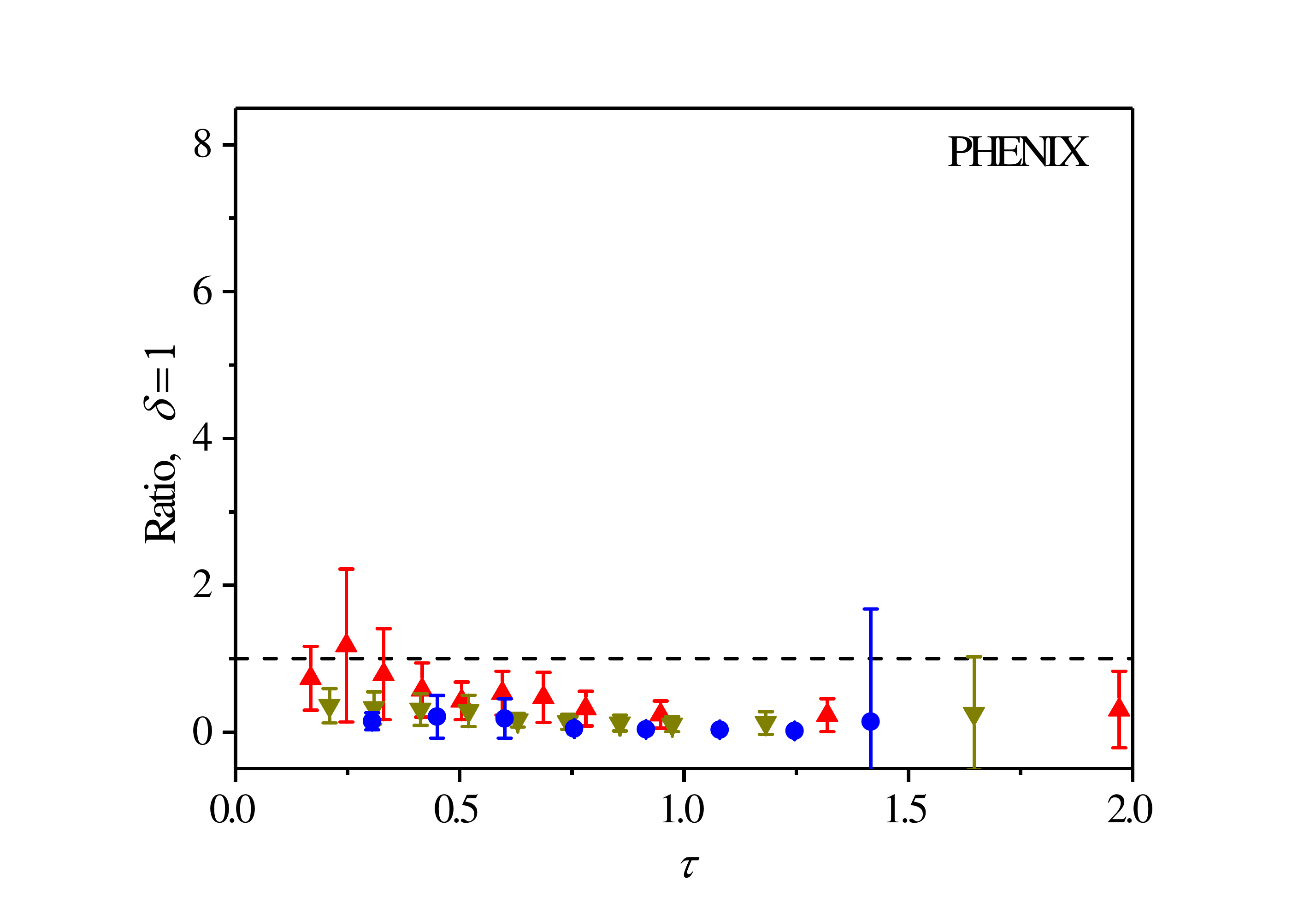}
\caption{Illustration of geometrical scaling of the $\gamma$ yields in heavy ion collisions at
different centrality classes. Left panel: ALICE. Red triangles correspond to $R_{c_1/c_2}$ and
dark-grren ones to $R_{c_1/c_3}$. 
Right panel: PHENIX. Red triangles points correspond to $R_{c_1/c_2}$,
dark-green ones to $R_{c_1/c_3}$ and blue ones to  $R_{c_1/c_4}$. Upper plot $\delta=1/3$, middle plot $\delta=2/3$ and the lower plot $\delta=1$.}
\label{fig:alph_ratio}
\end{figure}

In order to examine the quality of $N_{\rm part}$ scaling we construct the ratios of the scaled spectra at different
centralities $c_1$ and $c_2$ , $c_3$ or $c_4$
\begin{equation}
R_{c_1/c_{2,3}}(\tau)=\frac{1}{N_{1\;\text{part}}^{\delta}}\frac
{dN^{(1)}_{\gamma}}{N_{\text{evt}} \, 2\pi p_{\text{T}}\,d\eta dp_{\text{T}}}(\tau) \bigg/
\frac{1}{N_{2,3\;\text{part}}^{\delta}}\frac
{dN^{(2,3)}_{\gamma}}{N_{\text{evt}} \, 2\pi p_{\text{T}}\,d\eta dp_{\text{T}}}(\tau)
\label{eq:ratios}
\end{equation}
and plot them for different $\delta$ as functions $\tau$ in  Fig.~\ref{fig:alph_ratio}. In the left panel we plot
ALICE data where red up-triangles points correspond to 
$c_2=20-40\%$ and the dark green down-triangles to $c_3=40-80\%$. Geometrical scaling is achieved when 
$R_{c_1/c_2}\approx  R_{c_1/c_3}\sim 1$. As one can see
from Fig.~\ref{fig:alph_ratio}, this happens indeed for $\delta \approx 2/3$. The same ratios for PHENIX data are plotted in the right
panel of Fig.~\ref{fig:alph_ratio} where red triangles  correspond to $c_2=20-40\%$,
dark-green ones to $c_3=40-60\%$ and blue ones to  $c_4=60-92\%$.

\subsection{Energy scaling}

\begin{figure}[h!]
\centering
\includegraphics[width=6.8cm]{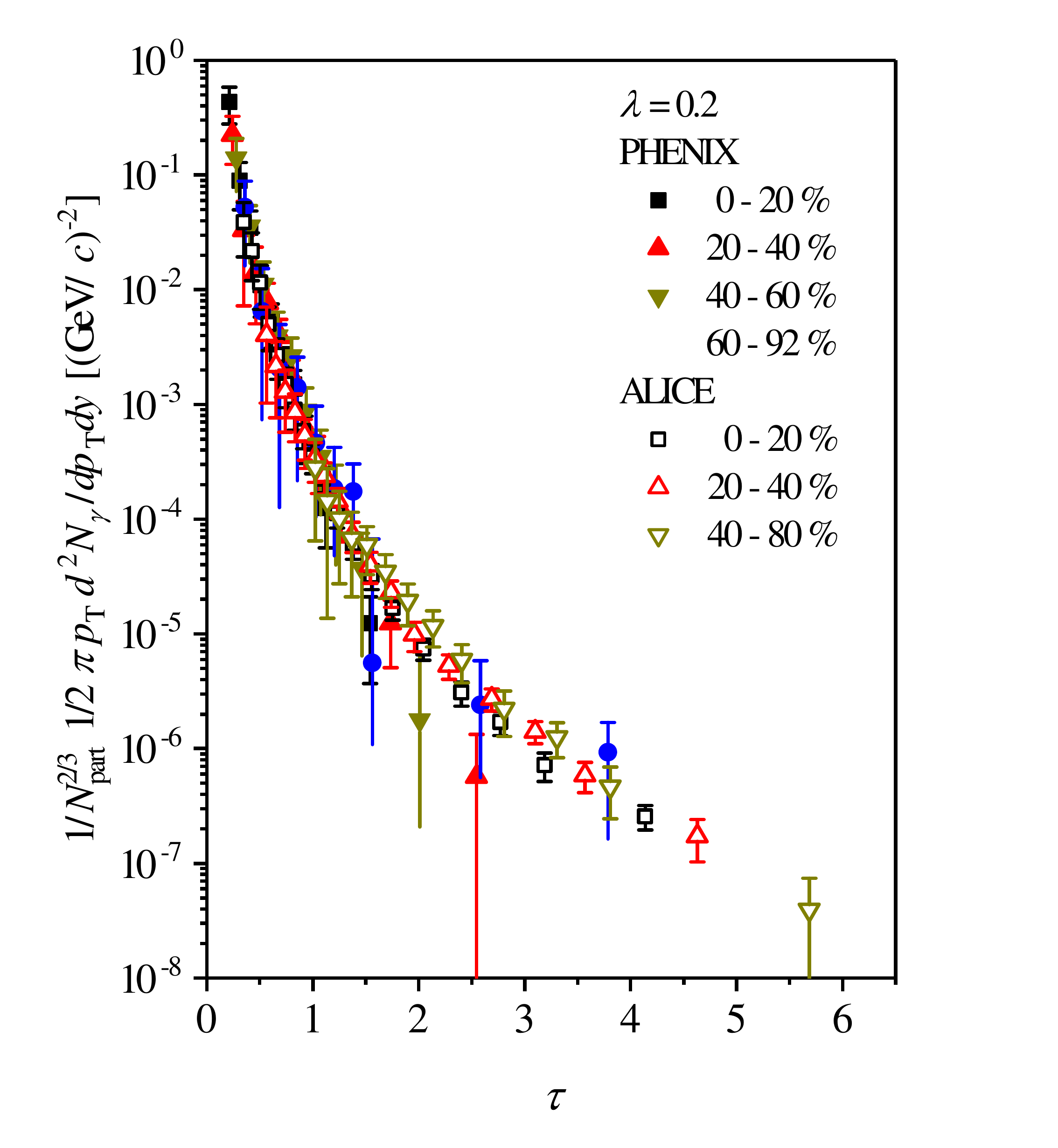}
\includegraphics[width=6.8cm]{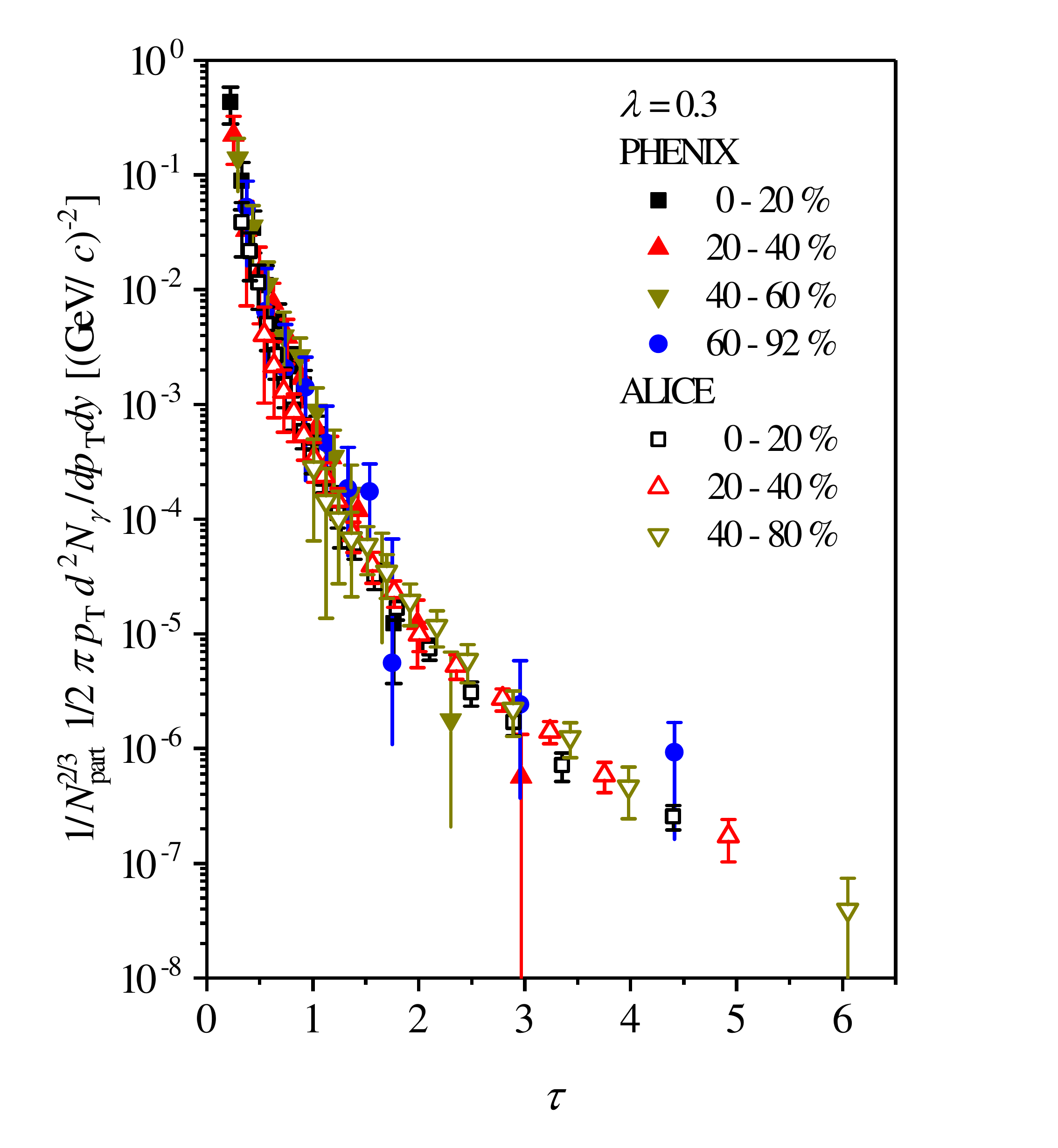}
\caption{Illustration of  geometrical scaling of the $\gamma$ yields in heavy ion collisions for
different centrality classes for PHENIX AuAu data at 200~GeV~\cite{Adare:2014fwh} and Alice data at 2.76~TeV~\cite{Adam:2015lda}. 
Left panel: $\lambda=0.2$, right panel $\lambda=0.3$.
Lower error bars without end caps have been arbitrarily shortened for better visibility.}%
\label{fig:Enscaling}
\end{figure}

Having established that GS is indeed achieved for $\delta = 2/3$ (or very close to 2/3), we can now test energy scaling,
{\em i.e.} $\lambda$ dependence of GS. Unfortunately, the quality of ratios, similar to the ones defined in Eq.~(\ref{eq:ratios}),
but for different energies rather than centralities, is very poor as compared to (\ref{eq:ratios}). This is because the data from different experiments
suffer from systematic uncertainties, like different rapidity ranges, different definition of centrality classes, etc. Therefore in
Fig.~\ref{fig:Enscaling} we simply plot spectra both for ALICE and PHENIX in terms of scaling variable  $\tau$ defined in Eq.~(\ref{tauHI})
for two different choices of $\lambda$: 0.2 in left panel and 0.3 in right panel. We see that it is hard to decide for which $\lambda$ GS
is better (with logarithmic accuracy). This shows that it is of importance to have data at different energies (at least three) from one experiment 
where the systematic uncertainties mentioned above cancel.

\section{Conclusions}
\label{concl}

Geometrical scaling is the property of the overoccupied gluonic cloud that is characterised by the new dynamical scale, called
{\em saturation scale}. If this is the only energy scale in a given kinematical region, then by a simple argument of dimensional analysis
particle spectra should depend only on the ratio of the transverse momentum to this scale. This phenomenon has been first
observed in inclusive DIS and then also in hadronic collisions. In the latter case the emergence of GS is by no means obvious due
the the final state interactions, resonance decays, confinement, etc. Nevertheless GS is observed in pp scattering and also in HI collisions.

Photons that have weak final state interactions (they are rather insensitive to the QGP medium) and are free from the confinement effects
seem to be a much better probe of the initial state than charged particles. This is, however, not entirely true, since photons do not
couple to gluons, and therefore the fact that they still exhibit GS provides a positive test of quark production mechanism in glasma.

The analysis presented here should be carried out on the recent PHENIX data~\cite{Adare:2018wgc} including also the data obtained from the collisions of different systems (Cu+Cu \cite{Adare:2018jsz}, Au+Au \cite{Afanasiev:2012dg}, Pb+Pb) and small systems (p+p, d+Au, p+Au)
-- for review see Ref.~\cite{Khachatryan:2018evz}.

\section*{Ackonwledgenments}
The author acknowledges very useful discussions with Larry McLerran and Vladimir Khachatryan. This research has been supported 
the Polish National Science Centre grant 2014/13/B/ST2/02486.

%
%
%

\end{document}